\documentstyle[aps,prl,multicol,epsf]{revtex}
\begin{document}
\widetext
\title{Transport Spectroscopy of Single-Walled Carbon Nanotubes}
\author{David H. Cobden, Marc Bockrath, Nasreen Chopra, Alex Zettl 
and Paul L. McEuen}
\address{Department of Physics, University of California at Berkeley, and
Lawrence Berkeley National Laboratory MS 2-200, Berkeley, CA94720}
\author{Andrew Rinzler, Andreas Thess and Richard E. Smalley}
\address{Center for Nanoscale Science and Technology, Rice Quantum 
Institute, and Departments of Chemistry and Physics, Mail Stop 100, Rice 
University, P.O. Box 1892, Houston, TX 77251}

\date{\today, e-mail: dcobden@physics.berkeley.edu}\maketitle

\begin{abstract} 
We have performed transport spectroscopy on individual
ropes of single-walled carbon nanotubes.  We find that the levels are
Zeeman split in a magnetic field, with a g-factor of $2.04 \pm 0.05$.  The
observed pattern of peak splittings indicates the parit y of the number of
electrons on the dot.  In one device there are also signs of the presence
of a second dot.  We observe features which resemble anticrossings between
quantum levels in the two dots, which may be formed from separate
conducting nanotubes within the rope.  
\end{abstract}

\begin{multicols}{2}
\narrowtext

In recent transport measurements on individual single-walled carbon
nanotubes \cite{tans97} and bundles, or "ropes", of single-walled carbon
nanotubes \cite{bockrath97}, evidence was found that the electrons were
confined to one-dimensional (1D) channels \cite{hamada92} of a finite
length.  The devices behaved as quantum dots, and it was therefore
possible to use nonlinear current-voltage ($I$-$V$) measurements to reveal
excited states, whose spacing was found to be consistent with expectations
for partic les in a 1D box.  Also, in a magnetic field Tans \textit{et
al.} \cite{tans97} observed an excited state which moved relative to the
ground state at a rate corresponding to a g-factor of $1.9 \pm 0.2$,
consistent with the expected free-electron Zeeman shi ft.  A surprising
aspect of their data was the apparent absence of the expected splitting of
the ground state transition. 

Here we present results of detailed transport spectroscopy (measurement of
$dI/dV$ as a function of bias $V$ and gate voltage $V_g$) at millikelvin
temperatures on a short rope segment.  We find that a magnetic field
produces a Zeeman splitting with a g-f actor of $2.04 \pm 0.05$.  The
pattern of splittings of the peaks in $dI/dV$ allows us to assign even or
odd parity to the number of electrons on the dot \cite{ralph95} over a
range of more than ten consecutive Coulomb peaks.  We also find evidence
of the presence of a second weakly conducting dot, and observe apparent
anticrossings of the levels in two dots.  The two dots may be formed from
two coupled conducting nanotubes within the rope. 

To make a device \cite{bockrath97} we deposit the nanotube material
\cite{thess96} sparsely on SiO$_2$, locate an isolated rope relative to
prefabricated alignment marks using an atomic force microscope (AFM), and
deposit chrome-gold contacts over it by e lectron-beam lithography.  In an
improvement on our earlier technique we use a degenerately doped silicon
substrate to act as a metallic gate.  An image of a device is inset to
Figure 1.  The diameter of the rope is about 5 nm, so that it consists of
arou nd ten nanotubes.  Only one segment of this rope, to which source and
drain leads are shown schematically attached, conducted. 

Figure 1 shows the small-signal conductance, $G=dI/dV|_{V=0}$, at 100 mK,
as function of gate voltage $V_g$ applied to the conducting substrate.  It
exhibits a series of sharp peaks of varying height and spacing.  If
measurements are limited to a range of about 2 V in $V_g$ these peaks are
highly reproducible.  The results of applying a dc bias $V$ are best
represented in a grayscale plot of $dI/dV$.  Figure 2 shows such a plot
for the region of $V_g$ around the three peaks labeled P, Q and R in
Figure 1.  Each dark line in this figure is the locus of a peak in
$dI/dV$.  Each of the peaks at $V = 0$ becomes a cross in the $V-V_g$
plane, with extra structure inside it which is different for each cross.

The effect of a magnetic field $B$ parallel to the rope is shown in Figure
3 (a) for the cross formed by peak Q.  Most of the lines, such as those
marked X and Y here, split into pairs (X splits into X$_1$ and X$_2$)
whose separation is proportional to $B $.  Some lines however (such as
line Z) do not split.  In the remainder of the data we observe the
following pattern: on successive crosses alternately the leftmost or the
rightmost line does not split.  The splitting is the only discernible
effect of $B$ on the positions of any of the lines.

As discussed previously \cite{tans97,bockrath97}, the peaks in $G$ in
Figure 1 can be explained by the Coulomb blockade model of a quantum dot
\cite{kouwenhoven97}.  In this model peaks in $dI/dV$ (dark lines in
Figure 2) are attributed to the alignment o f quantum levels in the dot
with the Fermi levels in the contacts.  The sketches in Figure 3 (b)
illustrate this with regard to the $B=0$ grayscale in Figure 3 (a).  Each
level is indicated by two horizontal lines to represent the spin
degeneracy.  The bi as V is applied to the left contact with the right
contact grounded.  We take the number of electrons $N$ on the dot to the
left of cross Q to be even, for reasons given below.  The lowest available
level for $N+1$ electrons is doubly degenerate, and lie s a distance $U+
\delta E$ above the highest filled level for $N$ electrons, where $U$ is
the charging energy and $\delta E$ is the level spacing.  Along line X
this level aligns with the Fermi level of the source, while along line Z
it aligns with that o f the drain.  Line Y is associated with alignment of
the source with an excited level, as indicated in Figure 3 (b).  The
difference in $V_g$ between X and Z at any bias $V$ produces a change of
$V$ in the electrostatic po-tential of the dot.  From this w e deduce that
a change in $V_g$ should be multiplied by $\alpha = 0.09$ to obtain the
corresponding change in dot potential. 

From the typical separation of lines like Y and X we estimate the average
level spacing to be $\delta E \sim 5$ meV, while equating the average
spacing in $V_g$ of the peaks in $G$ with $(U+ \delta E)/e \alpha$ we
obtain $U \sim 25$ meV.  We may compare these parameters with the expected
properties of a quantum dot formed from a single nanotube.  Assuming the
nanotube acts as a 1D box for the electrons, the average level spacing
should be $\delta E = \pi \hbar v_F/2L = (0.5$ eV nm$)/L$, where $v_F =
4.8 \times 10^5 $ms$^{-1}$ \cite{kwon97} is the Fermi velocity, $L$ is the
length and the 2 accounts for the two 1D bands at the Fermi energy.  Also,
the charging energy of a small object of size $L$ incorporated into an
electronic device is typically $U \sim (1.5$ eV nm$)/L$.  For $L = 100$
nm, these estimates yield $\delta E = 5$ meV and $U \sim 15$ meV.  We
conclude that a nanotube of length approximately equal to the distance
between the contacts (in this case lithographically defined as 200 nm)
should have both a level spacing and a charging energy similar to the
measured values.  The same was true in our previous rope devices
\cite{bockrath97}, while in the work of Tans \textit{et al.} \cite{tans97}
the level spacing and charging energy were consistent with $L = 3$ $\mu$m,
the entire length of the nanotube. 

Most ropes consist of tubes with a range of chiralities
\cite{thess96,luchang97}, and the fraction of metallic tubes may be small
\cite{dekker97,lieber97}.  Hence the origin of the dot in our device is
likely to be a single conducting nanotube within the rope, bounded
lengthwise by the metal contacts.  Barriers at the contacts may be created
by contamination or damage during fabrication, or by interaction between
the contact metal and the tube \cite{delaney97}. 

The effect of magnetic field in Figure 3 (a) can be understood within the
Coulomb blockade model as resulting from the Zeeman splitting.  The fact
that the only effect of $B$ is to split some of the lines is consistent
with the absence of orbital coupling to a magnetic field expected for a 1D
conductor.  Figure 3 (c) indicates the lifting of the spin degeneracy in
Figure 3 (b) by the Zeeman energy $g \mu_B B$.  With $N$ even, the
$N+1$'th electron must tunnel from the source into the next available
orbit al level.  For $B > 0$ the spin-down state aligns with the source
along line X$_1$ and the spin-up state along X$_2$.  The separation of
X$_1$ and X$_2$ in $V_g$ is $g \mu_B B/ \alpha$.  From a series of
measurements at different $B$ we obtain $g = 2.04 \ pm 0.05$, in agreement
with the g-factor of $2.001 \pm 0.001$ yielded by electron spin resonance
measurements on bulk nanotube material \cite{charlier95}. 

The line marked Z on the right of the cross, which does not split,
corresponds to electrons tunneling out of the spin-down $N+1$'th level to
Fermi level in the drain.  In this case the spin-up state produces no
line, because once the lower-energy spin-dow n state is below the drain
Fermi level it is permanently occupied, excluding occupancy of the spin-up
state.  Similar arguments show that if $N$ is odd, the opposite pattern is
seen: the line on the right of the cross splits while the one on the left
does not \cite{ralph95}.  We find that the pattern of splittings
alternates between these two types over at least ten crosses.  This
implies that the electrons are added sequentially to the dot just as
expected in the single particle picture, first spin d own then spin up for
each level. 

The picture described earlier of a rope as a collection of mostly
insulating nanotubes with an occasional metallic one helps us interpret
some observed deviations from the single-dot Coulomb blockade model.
Figure 4 shows grayscale plots covering peaks G , H, K, L and M in Figure
1.  Faint additional lines can be seen in between the crosses, moving at a
shallow angle to the $V_g$ axis.  These extra lines are associated with an
indistinct cross (not shown) interstitial between peaks I and J.  This
other cr oss indicates the presence of a second dot, which we can deduce
is poorly coupled to at least one of the contacts.  From a closer analysis
we can estimate the interdot charging energy to be $\sim 2$ meV.  The
second dot has a smaller capacitance to the ga te than the first so its
levels move at a different rate with $V_g$.  If the dots are
tunnel-coupled the states in them should hybridize and the resulting
levels anticross.  We suggest that the anticrossings apparent in Figure 4
are of this nature, and we can estimate a coupling of $\sim 1$ meV from
the minimum separation at the anticrossing.  It is possible that the
second dot is a different conducting nanotube within the rope.  Since the
coupling between adjacent tubes is expected to be large \cite{kwon97},
the two tubes may not be in direct contact.  Finally we remark that a
variety of aspects of the data, some of which are evident on close
inspection of Figure 4, remain mysterious and hint at exciting discoveries
to come in this novel system. 

In conclusion, we have performed detailed transport spectroscopy on an
individual rope of single-walled carbon nanotubes.  We observe a level
spacing and Zeeman splitting consistent with a single-particle model of a
finite 1D conductor, and see possible e vidence for tunnel coupling
between different conducting nanotubes within the rope.  

\acknowledgements
We would like to acknowledge useful discussions with
Cees Dekker, Sander Tans and Steven Louie.  The work at LBNL was supported
by the Office of Naval Research, Order No. N00014-95-F-0099 through the
U.S. Department of Energy under Contract No. DE-AC03-76SF00098, and by the
Packard Foundation.  The work at Rice was funded in part by the National
Science Foundation and the Robert A. Welch Foundation.

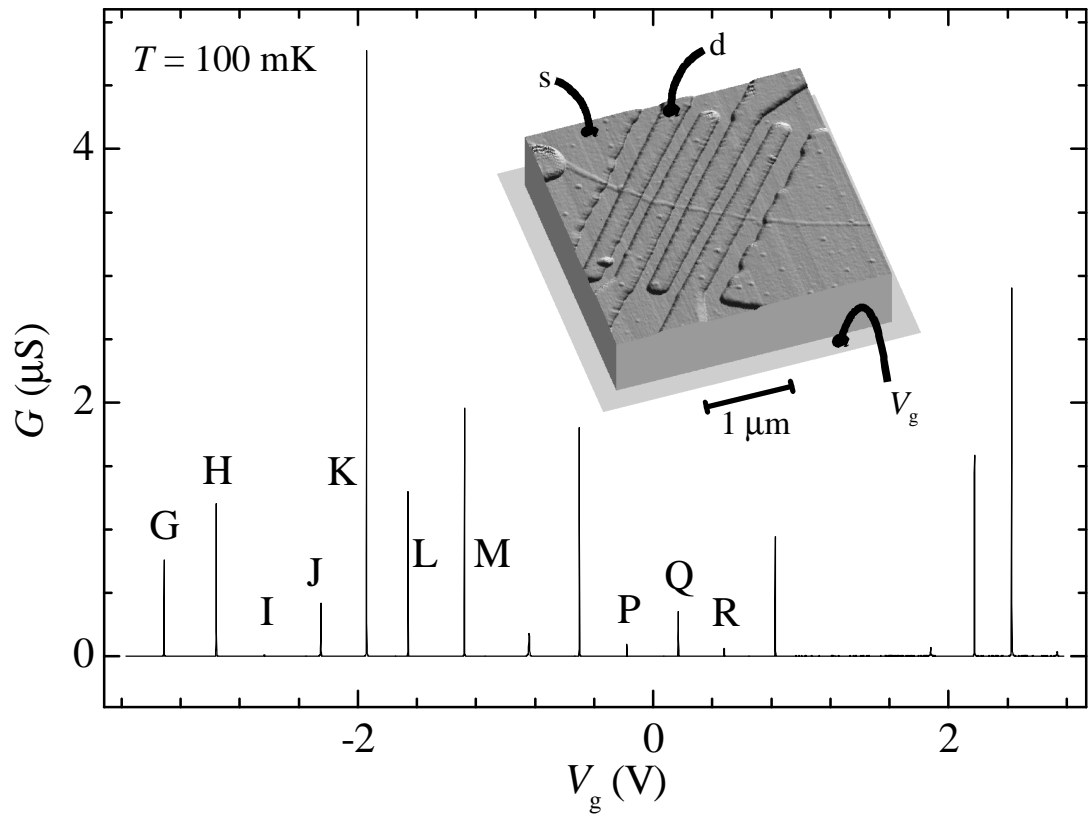
\begin{figure}
\caption{Small-signal two-terminal conductance $G$
of a single-walled nanotube rope vs gate voltage $V_g$.  Inset: AFM image
indicating the leads used in the measurement.}
\label{fig1}
\end{figure}

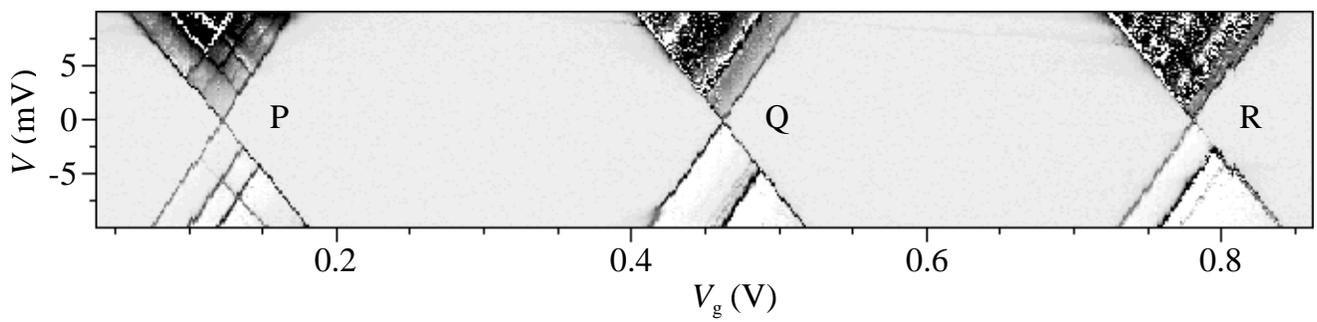
\begin{figure}
\caption{Grayscale plot of $dI/dV$ (dark = more positive) vs bias $V$ 
and $V_g$ for peaks P, Q and R in Figure 1.}
\label{fig2}
\end{figure}

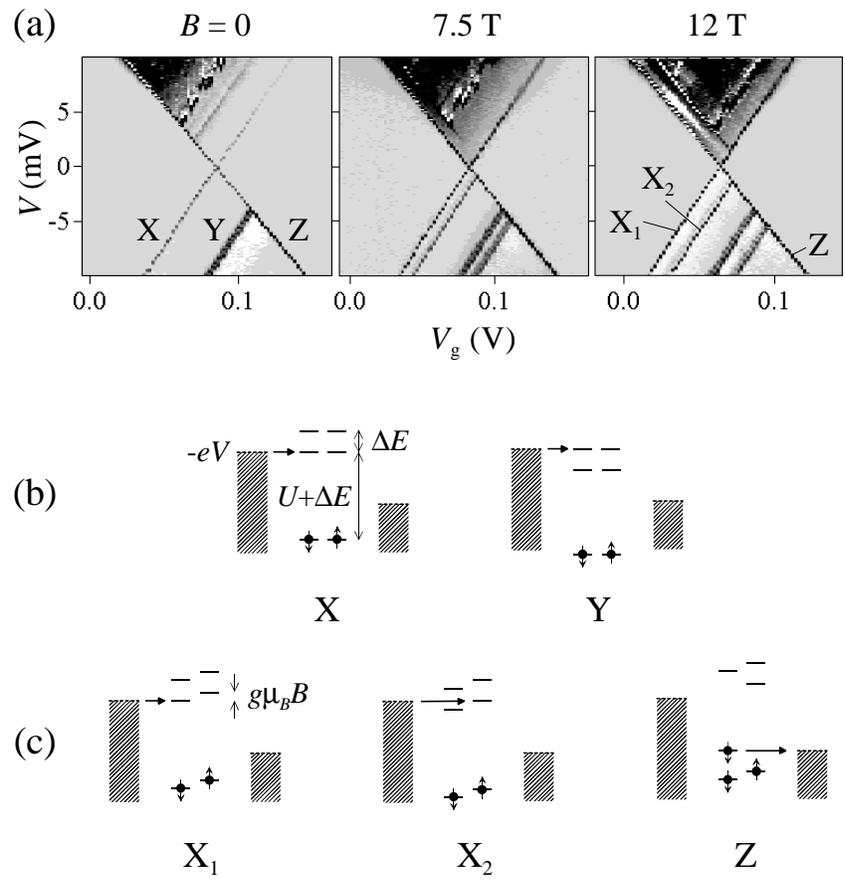
\begin{figure} 
\caption{(a) Grayscale plot of $dI/dV$ for peak Q
in Figure 2 at a series of magnetic fields.  (b) Coulomb-blockade model
for the features (dark lines) in (a) at $B = 0$.  (c) Effect of Zeeman
splitting on the features at $B > 0$.}
\label{fig3}
\end{figure}

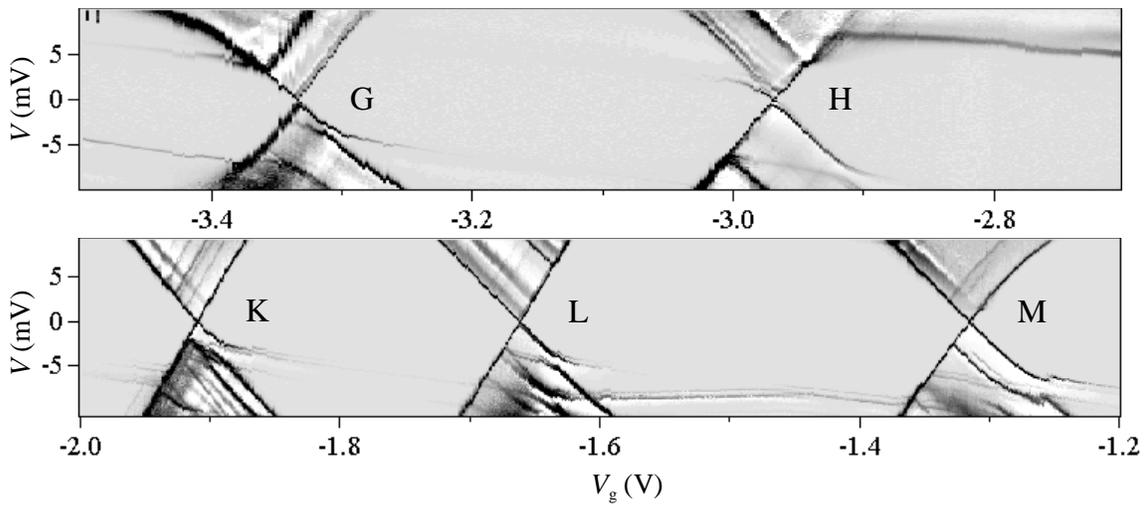
\begin{figure}
\caption{Grayscale plots of $dI/dV$ for peaks G, H,
K, L and M in Figure 1, showing anticrossings and other distortions
possibly related to the presence of a second dot.}
\label{fig4}
\end{figure}

\end{multicols}
\widetext

\end{document}